# The Bones and Shapes of the Phillips Curve

Hanyuan Jiang*

London School of Economics and Political Science

**Abstract**

The COVID-19 pandemic reignited debate on the U.S. Phillips curve. Using MSA-level panel data (2001-2024), we employ a Two-Stage Least Squares (2SLS) instrumental variable strategy with a shift-share instrument to estimate core non-tradable inflation's response to a v/u-based slack measure. We distinguish structural slope ($\kappa$) stability from state-dependent non-linearities ($\lambda(\theta)$) via a threshold model. Our analysis addresses whether the slope of the Phillips Curve changed during and after the Pandemic in the United States by evaluating if recent inflation dynamics reflect an altered structural trade-off ("bones") or the activation of non-linear "shapes" in response to extreme labor market tightness. This distinction offers critical insights into the unemployment cost of disinflation.

JEL Classification: E30

* I am grateful to Emi Nakamura, Jonathon Hazell, Maarten De Ridder, and Marie Fuchs for helpful guidance, discussions, and suggestions, and to participants at workshops at LSE for comments and discussions. The views expressed herein are those of the author(s); all errors remain our own.

# 1. Introduction

The Pandemic rekindled the old debate: does the U.S. Phillips curve flatten, steepen, or simply shift when the economy traverses extreme states of slack? Aggregate time-series evidence is notoriously fragile as movements in long-run inflation expectations contaminate slope estimates, implying a near-flat slope over the past two decades, yet the juxtaposition of low unemployment and rapid price growth appears at odds with that conclusion. Resolving this tension matters for policy, if the pandemic shifted the structural inflation–slack trade-off, the unemployment cost of disinflation might be materially higher than pre-pandemic estimates suggest; if, instead, the slope is state-dependent but the underlying "bones" of the curve are unchanged, a modest cooling of labor-market tightness could suffice to anchor prices. Our research asks whether the slope of the Phillips curve in the United States changed during and after the pandemic and, crucially, whether any change reflects a permanent alteration of structural parameters or a transitory, regime-dependent amplification.

Our research builds upon two literatures. First, we extend the work of Hazell et al. (2022), who exploit cross-state variation in non-tradable prices to show that the structural slope, $\kappa$, was already low by the early 1980s and remained remarkably stable through 2019. Their identification strategy, which eliminates national movements in inflation expectations by differencing out time fixed effects, provides a powerful framework for isolating the purely local covariation between unemployment and inflation. We adopt this cross-sectional logic, extending their sample to 2024 to cover the pandemic and its aftermath. Second, we confront the possibility, highlighted by Benigno and Eggertsson (2023), that the effective slope, $\lambda$, may rise sharply when the vacancy-to-unemployment ratio, $\theta \equiv v/u$, exceeds a Beveridge threshold of roughly one. Their non-linear New-Keynesian framework attributes the recent inflation surge to a state-contingent jump in marginal hiring costs once vacancies outstrip job seekers. Our research design explicitly aims to distinguish the stability of the structural 'bones' ($\kappa$) from the potential state dependence of the effective 'shape' ($\lambda(\theta)$).

The broader context for our study is the long-standing debate on the Phillips curve's apparent 'flattening' since the 1990s, which led some to question its policy relevance (e.g., Blanchard, 2016). A prominent explanation for this phenomenon centered on the role of anchored inflation expectations, as argued by Bernanke (2007) in the context of the Great Moderation. Coibion and Gorodnichenko (2015) further refined this by showing that imperfectly anchored or heterogeneous expectations, particularly those of households and firms, could explain anomalies like the 'missing disinflation' post-Great Recession. More recently, Reis (2021) emphasized information rigidities, suggesting that agents' slow updating of information sets further complicates observed inflation dynamics and Phillips curve estimations. These perspectives underscore the challenges in using aggregate data to identify a stable inflation-slack trade-off when expectation dynamics are complex and time-varying. The cross-sectional approach, by differencing out common national expectation shocks, offers a path to a clearer estimate of the underlying structural slope, as demonstrated by Hazell et al. (2022).

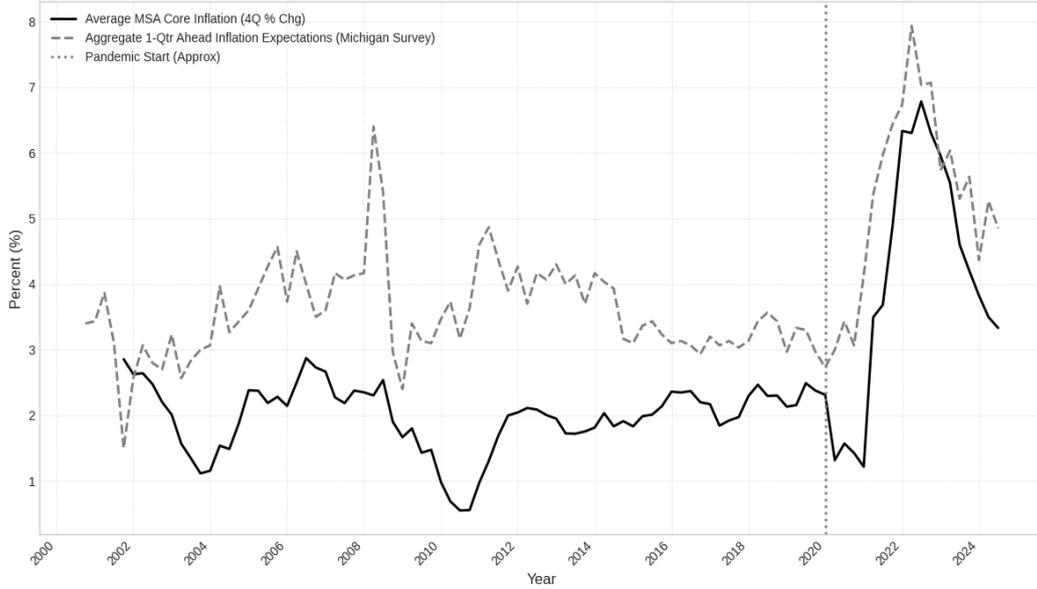

FIGURE I

Core Inflation and Short-Term Inflation Expectations

Figure I illustrates the central identification challenge our study addresses. The figure plots average core non-tradable inflation across MSAs alongside aggregate short-term inflation expectations from 2000 to 2024. Two features stand out. First, during the 2008-2012 period, core inflation plummeted to near zero while expectations remained elevated around 4 percent, the quintessential "missing disinflation" that confounded Phillips curve estimates. This disconnect between realized inflation and expectations persisted through much of the 2010s, with expectations consistently overshooting actual inflation by 1-2 percentage points. Second, and most strikingly, the pandemic period marks a regime shift: both series surge in tandem, with core inflation reaching 7 percent and expectations approaching 8 percent by 2022. This co-movement suggests that the inflation dynamics of 2021-2023 differ fundamentally from the prior two decades, either because expectations became unanchored, reminiscent of the 1970s, or because an underlying nonlinearity in the Phillips curve was activated by extreme labor market tightness. The time-series correlation between these variables makes it virtually impossible to separately identify the slope of the Phillips curve from expectation effects using aggregate data alone, motivating our cross-sectional approach that exploits variation across MSAs while controlling for common national movements in expectations through time fixed effects.

To empirically address these questions, we leverage the quarterly panel dataset at the U.S. Metropolitan Statistical Area (MSA) level from 2001 to 2024. Our analysis centers on core non-tradable Inflation, constructed from MSA-specific CPI data and a measure of labor market slack derived from the vacancy-to-unemployment ratio ($v/u$), which is particularly suited for capturing labor market tightness in extreme states. Methodologically, we employ an instrumental variable strategy using a shift-share instrument equivalent to that in Hazell et al. (2022) to identify the structural Phillips curve slope ($\kappa$). We then explore non-linearities by examining how this slope varies with the level of labor market tightness ($\theta \equiv v/u$). This dual investigation allows us to investigate whether the perceived changes during the pandemic reflect a shift in the fundamental 'bones' of the Phillips curve or an activation of its latent non-linear 'shape'. The subsequent section will establish the theoretical basis for these two avenues of inquiry, connecting micro-level price-setting behavior (relevant for $\kappa$) with macroeconomic models of state-dependent slope coefficients (relevant for $\lambda(\theta)$).

## 2. Foundations: Calvo Pricing and Slack-Driven Nonlinearity

Our empirical strategy rests on two intertwined pillars, the mechanisms that determine its structural slope and its potential state-dependent non-linearities. This section discusses the relevant contributions and clarifies how each informs the specifications that follow.

### 2.1 The New Keynesian Phillips Curve and the Structural Slope

The New Keynesian Phillips Curve (NKPC) emerges from microfoundations incorporating nominal rigidities, typically through the Calvo (1983) model of staggered price setting. In this framework, firms are assumed to adjust their prices infrequently. Each period, only a fraction $(1 - \xi)$ of randomly selected firms can set a new optimal price ($p_t^*$), while the remaining fraction ($\xi$) keeps their prices unchanged. This sluggish price adjustment at the firm level aggregates up to a relationship between current inflation ($\pi_t$), expected future inflation ($E_t[\pi_{t+1}]$), and a measure of real marginal cost ($m_{ct}$). The canonical log-linearized NKPC can be expressed as:

$$\pi_t = \beta E_t[\pi_t + 1] + \kappa m_{ct},$$

where $\beta$ is the firm's discount factor. The structural slope parameter, $\kappa$, quantifies the responsiveness of inflation to real marginal costs and is fundamentally determined by the degree of price stickiness and other structural parameters of the economy. Specifically, in a basic Calvo framework, $\kappa$ is given by:

$$\kappa = \frac{(1 - \xi)(1 - \beta \xi)}{\xi}.$$

This expression highlights that a higher degree of price stickiness (a larger $\xi$) leads to a smaller $\kappa$, implying a flatter Phillips curve. Seminal work by Nakamura and Steinsson (2008), re-evaluating price-setting behavior even outside of sales, found that U.S. consumer prices reset roughly every four to five quarters. This implies a Calvo non-adjustment probability $\xi$ of approximately 0.75. Subsequent research using scanner data, such as Eichenbaum, Jaimovich & Rebelo (2011), has provided broadly similar estimates of price-setting frequency. The documented stability of these price-setting frequencies over decades, as also noted by Hazell et al. (2022) for non-tradables, underpins the rationale for testing whether this structural slope $\kappa$ has remained stable through the recent pandemic period. While more elaborate models derive $\kappa$ as a function of deeper structural parameters (like the elasticity of labor supply or the intertemporal elasticity of substitution in consumption), our empirical approach will focus on the reduced-form implications and the mapping to $\kappa$ via estimated persistence.

## 2.2 Search and Matching Frictions

Recent works, particularly by Benigno and Eggertsson (2023, 2024), have re-formalized Phillips' (1958) insight of a non-linear relationship between inflation and labor market conditions, nesting it within a New Keynesian framework. They propose a "Slanted-L Phillips curve" where the economy's response to shocks dramatically changes with labor market tightness, measured by the vacancy-to-unemployment ratio, $\theta \equiv v/u$. This measure of economic slack is standard in the search and matching literature (e.g., Mortensen and Pissarides, 1994; Michaillat and Saez, 2023). Their model incorporates search and matching frictions and a specific form of wage rigidity to generate these strong non-linearities.

The core mechanism hinges on how firms' marginal costs, particularly the cost of hiring new workers, behave in different labor market states. In their model, when the labor market is slack ($\theta < \theta^*$, where $\theta^*$ is a threshold (Beveridge threshold) often approximated by $\theta^* = 1$), wages may fall only slowly, consistent with Phillips' (1958) observation about workers' reluctance to accept wages below prevailing rates. However, when the labor market becomes exceptionally tight ($\theta > \theta^*$, a labor shortage where vacancies outstrip job seekers), firms are "tempted to outbid one another", and the cost of hiring new workers becomes the more relevant measure of marginal costs for pricing decisions, rather than average wages. This occurs because when expanding production, firms primarily rely on attracting new workers. As empirical evidence from Crump et al. (2024) and others suggests, wages for new hires or posted wages can rise much more rapidly than average wages during such shortages. Benigno and Eggertsson (2023) formalize this by deriving a Phillips curve where the slope coefficient on labor market tightness ($\theta^t$) becomes significantly larger ($\kappa^{tight} > \kappa$) when $\theta_t > \theta_t^*$. Furthermore, their framework suggests that supply shocks ($\vartheta^t$) are also transmitted to inflation with greater force in a tight labor market ($\kappa_v^{tight} > \kappa_v$). This state-contingent jump in marginal costs and the amplification of supply shocks provide a theoretical basis for the " Slanted-L Phillips curve " and offer an explanation for the inflation surge in the 2020s, a period characterized by high $\theta$ values and anchored long-term inflation expectations, distinguishing it from the Great Inflation of the 1970s. Our empirical investigation of the "shape" of the Phillips curve directly tests this state-dependent slope, $\lambda(\theta)$, by allowing the semi-elasticity of inflation to slack to vary with $\theta_{it}$.

## 2.3 Instrumental Variable Strategy for Identification

Estimating the causal relationship between labor market slack ($s_{it}$) and inflation ($\pi_{it}$) is challenging due to the inherent endogeneity of slack. Unobserved aggregate or local demand and supply shocks can simultaneously affect both inflation and labor market conditions, leading to biased estimates of the Phillips curve slope if standard OLS is used. To address this, we employ an instrumental variable strategy centered on a Bartik-style (shift-share) instrument for generating plausibly exogenous variation in local economic conditions.

The Bartik instrument (Bartik 1991) leverages the principle that local economic outcomes are differentially affected by aggregate industry-level shocks based on the local area's initial industry composition. The instrument ($Z_{it}$) for local labor market slack in region $i$ at time $t$ is typically constructed as:

$$Z_{it} = \sum_k w_{ik,t_0} \times \Delta D_{kt},$$

where $w_{ik,t_0}$ represents the share of industry $k$ in region $i$ at a fixed initial period $t_0$ (e.g., pre-pandemic employment shares), and $\Delta D_{kt}$ is a measure of the national-level shock to industry $k$ at time $t$ (e.g., national log change in real spending or employment ). The intuition is that regions more specialized in industries experiencing positive national shocks will see a larger increase in local labor demand (and thus a decrease in slack), and vice versa. The exogeneity of the instrument relies on national industry shocks being uncorrelated with region-specific unobservables that affect local inflation, conditional on controls, and on the initial shares $w_{ik,t_0}$ being predetermined and uncorrelated with subsequent region-specific shocks. In our primary specification, we utilize a pre-constructed *shift_share* variable available in our MSA_q.csv dataset, which is described as a metropolitan area equivalent to that used by Hazell et al. (2022) and captures shocks to national demand for tradable industries.

While powerful, shift-share instruments have faced scrutiny regarding their validity and the interpretation of the estimates they produce. Key critiques, such as those highlighted by Goldsmith-Pinkham, Sorkin & Swift (2020), point to potential issues if the exogeneity of national shocks is violated or if specific industries (with large shares or large shocks) disproportionately drive the instrument's power. Adao, Kolesár & Morales (2019) also discuss challenges related to inference when error terms are correlated across regions with similar industry shares.

## 3. Data and Descriptive Statistics

In this section we detail the data used for the empirical analysis, the construction method of the core variables, and the basic descriptive statistics. Our analysis is based primarily on quarterly panel data at the US Metropolitan Statistical Area (MSA) level obtained from the MSA_q.csv file, with the sample period running from the first quarter of 2001 to the second quarter of 2024. These variables capture inflation dynamics at the local level as well as core features of the labor market to examine the structure and behavior of the Phillips curve before and after the pandemic.

### 3.1 Variables
#### 3.1.1 Non-Tradable Inflation ($\pi_{it}$)

Our measure of local inflation pressure focuses on core non-tradable goods and services. Consistent with Hazell et al. (2022), this approach minimizes contamination from nationally traded goods whose prices are largely determined in global markets and avoids idiosyncrasies related to shelter measurement, thereby providing a clearer signal of how local demand conditions influence price setting. We construct our non-tradable inflation measure, $\pi_{it}$, for each MSA $i$ in quarter $t$ using the $CPI\_core$ variable from the MSA_q.csv file. Specifically, inflation is defined as the four-quarter log difference of the core non-tradables CPI, annualized by multiplying by 100:

$$\pi_{it} = 100 \times [ln(CPI\_core_{it}) - ln(CPI\_core_{i,t-4})].$$

This four-quarter rolling window approach eliminates seasonality without resorting to dummy variable adjustments and aligns the inflation frequency with that of key labor market data, such as JOLTS vacancy statistics.

### 3.1.2 Labor Market Slack ($s_{it}$) and Tightness ($\theta_{it}$)

To capture varying conditions in the labor market, we construct two primary measures derived from the vacancy-to-unemployment ratio ($v/u$), which is available as the variable $vu$ in our MSA_q.csv dataset. The $v/u$ ratio is a crucial indicator of labor market conditions, especially in periods of extreme market states.

1. Labor Market Tightness ($\theta_{it}$). Following Benigno and Eggertsson (2023, 2024), we use the vacancy-to-unemployment ratio directly as our measure of labor market tightness $\theta_{it} = vu_{it}$. This variable is central to their non-linear New Keynesian framework and serves as the state variable in our non-linear Phillips curve specifications (Model II). An increase in $\theta_{it}$ signifies a tighter labor market, where firms' job vacancies are numerous relative to unemployed workers. Benigno and Eggertsson (2023) argue that when $\theta$ exceeds a Beveridge threshold (approximated by $\theta^* = 1$), marginal hiring costs can jump, leading to a steeper Phillips curve.

2. Labor Market Slack ($s_{it}$): For our linear Phillips curve specifications (Model I), we require a measure of labor market slack. While Hazell et al. (2022) primarily use the deviation of the local unemployment rate from its contemporaneous cross-sectional mean, we construct our primary slack indicator based on the inverse of labor market tightness, $1/\theta_{it}$ (which is $u/v$). This aligns with the focus on $v/u$ dynamics and allows for a consistent framework. To isolate local idiosyncratic variations, we then demean this measure by its contemporaneous cross-sectional quarterly mean:

$$S_{it} = (1/\theta_{it}) - meanq(1/\theta_{it'}).$$

A higher value of $S_{it}$ indicates greater slack in the local labor market relative to the national average tightness for that quarter. This demeaning process is important for our identification strategy as it distinguishes national movements in the natural rate of $v/u$ (or its inverse) and common shocks to expectations, following the logic applied by Hazell et al. (2022) to the unemployment gap. The choice of a $v/u$-based slack measure is further motivated by recent literature (e.g., Ball et al., 2022; Barnichon and Shapiro, 2022; Domash and Summers, 2022; Furman and Powell, 2021) suggesting that $v/u$ provides a better fit for inflation dynamics than traditional unemployment measures, especially during periods like the 2020s inflation surge when unemployment and $v/u$ conveyed different signals about labor market pressure.

### 3.1.3 Lagged Relative Price of Non-Tradables

Following Hazell et al. (2022), a key control variable in our specification is the lagged relative price of non-tradables. This variable is intended to capture potential mean-reversion in regional price levels. If a region's non-tradable prices were unusually high (or low) relative to a benchmark in the past, they might be expected to revert towards that benchmark, influencing current inflation dynamics independently of current labor market slack. In their empirical specification (eq.15), Hazell et al. (2022) include $\hat{p}_{it}^N$ as the percentage deviation of the home relative price of nontradeables from its steady-state value, and in equation (19) they use its fourth lag, $p_{i,t-4N}$. The theoretical underpinning for this variable arises from their multi-region model where the regional Phillips curve for nontradable goods inflation ($\pi_{Ht}^N$) explicitly includes a term for the relative price of non-tradables, $\lambda \hat{p}_{Ht}^N$ (eq.11). This term pushes relative prices toward parity in the long run.

To construct this variable for our MSA-level analysis, we use $CPI\_core$ (MSA-level core non-tradables CPI) and $CPI$ (MSA-level overall CPI) from our MSA_q.csv dataset. We define the contemporaneous relative price of non-tradables for MSA $i$ at quarter $t$ as

$$rp_{it} = ln(CPI\_core_{it}) - ln(CPI_{it}).$$

The control variable, $rel\_p\_lag$, is then the four-quarter lag of this measure:

$$rel\_p\_lag_{it} = rp_{i,t-4} = ln(CPI\_core_{i,t-4}) - ln(CPI_{i,t-4}).$$

### 3.2 Descriptive Statistics

Our primary estimation sample spans from 2001 Q1 to 2024 Q2. This period is determined by the availability of comprehensive MSA-level data for our variables, particularly the components needed for the vacancy-to-unemployment ratio ($vu$) that underpins our slack and tightness measures, and the pre-calculated $shift\_share$ instrument.

The summary statistics for the core variables are available in the Appendix. Non-tradable inflation ($\pi_{it}$, $pi\_core\_4q$) exhibits a full sample mean of approximately 2.51% with a standard deviation of 1.67. In the pre-pandemic period, average inflation was lower at 1.99%, rising markedly to an average of 4.03% during and after the pandemic, with a corresponding increase in volatility (standard deviation increasing from 1.00 to 2.22). Our primary measure of labor market slack ($s_{it}$, $slack$), by construction, has a mean of zero over the full sample. Its standard deviation is 0.68 in the full sample, decreasing from 0.74 pre-pandemic to 0.34 in the pandemic/post-pandemic period, suggesting a compression in the cross-MSA dispersion of relative slack, despite individual MSA experiences.

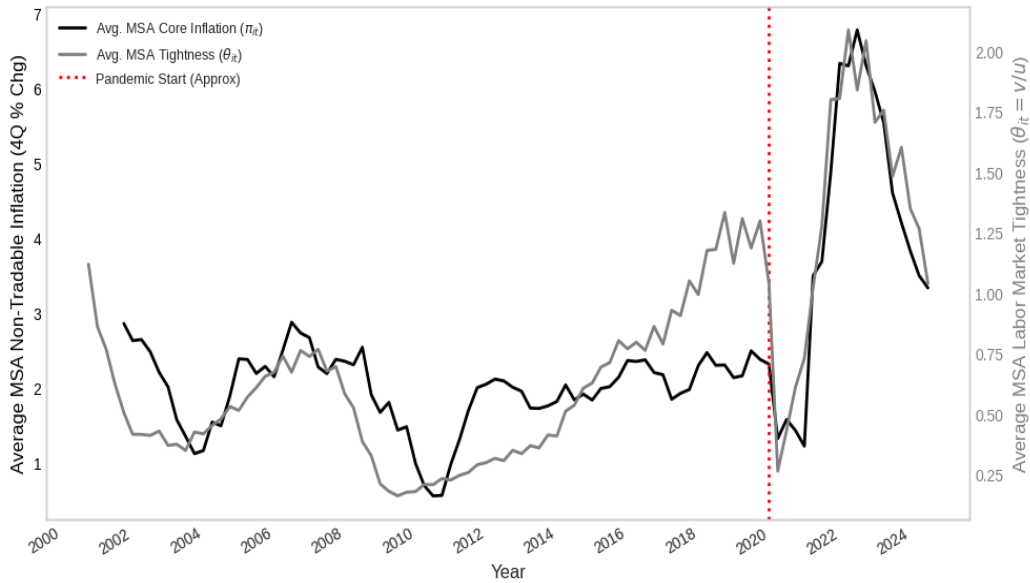

FIGURE II

Average MSA Non-Tradable Inflation and Labor Market Tightness

Labor market tightness ($\theta_{it}$, $theta\_it$ or $vu$) averaged 0.74 over the full sample. It was considerably lower in the pre-pandemic period (mean 0.59) and rose substantially to an average of 1.32 during and after the pandemic, with the maximum observed value reaching 3.48. This shift underscores the significant tightening of labor markets in the recent period. The instrumental variable ($Z_{it}$, $shift\_share$) shows a relatively stable mean around 0.018-0.019 across the full sample and pre-pandemic period, with a slight decrease to 0.012 in the pandemic/post-pandemic period, though its standard deviation increased. Finally, lagged relative non-tradable prices ($rel\_p\_lag$) maintain a small positive mean (around 0.018-0.020) across all periods, with relatively stable dispersion.

Figure II presents the time series evolution of our two key variables at the aggregate MSA level. The pre-Great Recession period (2000-2007) exhibits moderate inflation. Second, the Great Recession and recovery period (2008-2019) shows the dramatic collapse in both inflation and $\theta$, followed by a period of below-target inflation despite gradually tightening labor markets, which is the "missing inflation" puzzle that motivated much of the literature. Most strikingly, the pandemic period marks an unprecedented regime, $\theta$ rockets from below 0.5 to above 2.0, while inflation surges from near zero to almost 7%. This co-movement is particularly notable because $\theta$ crosses the theoretically significant threshold of 1.0 precisely when inflation begins its steep ascent.

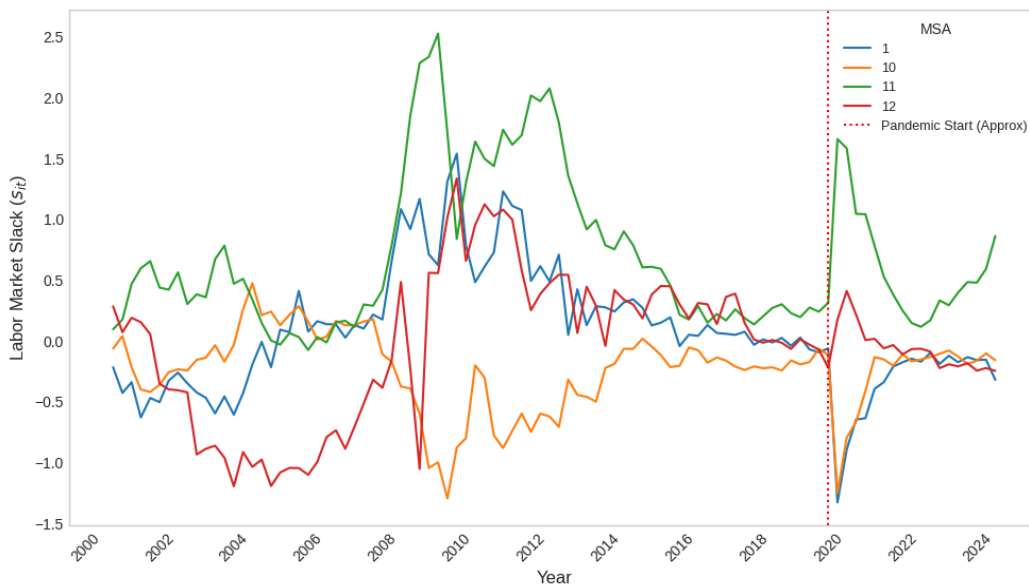

FIGURE III

Regional Labor Market Slack Heterogeneity

Figure III shows labor market slack for five representative MSAs over our sample period. The pandemic period exhibits even dramatic heterogeneity. MSA 11 experiences an extraordinary spike in slack exceeding 2.5 percentage points, likely reflecting an MSA heavily dependent on contact-intensive services, while other MSAs show more modest changes. Crucially, the post-pandemic recovery also varies substantially across MSAs, with some returning to pre-pandemic tightness by 2022 while others remain elevated. Moreover, the dramatic shift in average $\theta$ in Figure II provides the variation needed to test for state-dependent nonlinearities in the Phillips curve. The fact that this shift coincides with the inflation surge, while earlier periods of moderate $\theta$ movements saw little inflation response, somehow provides prima facie evidence for the threshold effects discussed by Benigno and Eggertsson (2023). In summary, the sample data reveals distinct patterns and significant shifts in both non-tradable inflation and labor market dynamics when comparing the pre-pandemic era to the Pandemic and its aftermath.

## 4. Empirical Strategy

We aim to distinguish between alterations in the underlying structural parameters (the "bones") of the Phillips curve and state-dependent variations in its effective slope (the "shape"). We first detail the model used to estimate the linear structural slope, $\kappa$, and test its stability. Subsequently, we will specify the model designed to explore non-linearities contingent on labor market tightness, $\lambda(\theta)$.

### 4.1 Model I: Estimating the Linear Phillips Curve Slope

Our primary approach to estimating the linear Phillips curve slope builds on the cross-sectional identification strategy by Hazell et al. (2022). Their methodology leverages regional (state-level in their case, MSA-level in ours) variation in non-tradable inflation and unemployment, using time fixed effects to difference out common national movements in long-run inflation expectations, thereby isolating local economic relationships. The key insight is that long-run inflation expectations ($E_t \pi_{t+\infty}^N$) are constant across regions in a monetary union and are thus absorbed by time fixed effects in a panel specification. Adapting this framework, we use our MSA-level measure of labor market slack, $s_{it}$, as the primary indicator of local economic activity. To test for changes in the slope across the pandemic, we estimate an interaction model using Two-Stage Least Squares (2SLS). The second-stage specification is:

$$\pi_{it} = \alpha_i + \delta_t + \psi_{pre} s_{it} + \Delta\psi(s_{it} \times \text{pandemic\_period}_{it}) + \phi \text{rel\_p\_lag}_{it} + \epsilon_{it}.$$

Here, $\pi_{it}$ is the 4-quarter core non-tradable inflation for MSA $i$ in quarter $t$. $\alpha_i$ represents MSA fixed effects, and $\delta_t$ captures time (quarter-year) fixed effects, which also absorb national long-run inflation expectations. $s_{it}$ is our measure of labor market slack. $\text{pandemic\_period}_{it}$ is a dummy variable equal to one for quarters from 2020 Q1 onwards, and zero otherwise. The coefficient $\psi_{pre}$ thus captures the Phillips curve slope in the pre-pandemic period, while $\Delta\psi$ measures the change in this slope during and after the pandemic. We include $\text{rel\_p\_lag}_{it}$, the four-quarter lagged relative price of non-tradables as a control variable, consistent with Hazell et al. (2022, eq. 19) who use beginning-of-period relative prices to control for mean-reversion in regional price levels. Due to the endogeneity of $s_{it}$ and its interaction term $s_{it} \times \text{pandemic\_period}_{it}$, we instrument them using $Z_{it}$ ($shift\_share$) and $Z_{it} \times \text{pandemic\_period}_{it}$, respectively.

From the estimated $\hat{\psi}_{pre}$ and $\Delta\hat{\psi}$, we derive the post-pandemic slope as $\hat{\psi}_{post} = \hat{\psi}_{pre} + \Delta\hat{\psi}$. To map these reduced-form slopes to the structural slope $\kappa$, we follow Hazell et al. (2022), who use the relationship $\psi = \frac{\kappa}{1-\beta\rho_u}$. The persistence of slack, $\rho_u$, is estimated from an AR(1) regression of $s_{it}$ on its lag, $s_{it} = c_i + \rho_u s_{i,t-1} + e_{it}$, including MSA fixed effects, separately for the pre-pandemic ($\hat{\rho}_{pre}$) and pandemic/post-pandemic ($\hat{\rho}_{post}$) periods. The discount factor $\beta$ is calibrated to $0.99^{1/4}$. This allows us to compute:

$$\kappa_{pre} = \hat{\psi}_{pre}(1 - \beta\hat{\rho}_{u,pre}),$$
$$\kappa_{post} = \hat{\psi}_{post}(1 - \beta\hat{\rho}_{u,post}).$$

Comparing $\kappa_{pre}$ and $\kappa_{post}$ will allow us to assess the stability of the Phillips curve's structural "bones" across the pandemic. We will ensure robust inference by clustering standard errors at the MSA level and paying close attention to first-stage F-statistics and weak instrument robust methods.

### 4.2 Model II: Estimating the State-Dependent Phillips Curve Slope

Beyond linear shifts, we also investigate its potential non-linearities, specifically whether its effective slope, $\lambda(\theta)$, is state-dependent, varying with the degree of labor market tightness. To empirically test for such state-dependence using our MSA-level panel data, we adopt a threshold linear interaction model. This simple approach captures the core inspiration of the non-linearity as discussed by Benigno and Eggertsson (2023). We define a dummy variable, $D_{it}^{\theta>\tau}$, which equals one if labor market tightness $\theta_{it}$ exceeds a pre-determined threshold $\tau$, and zero otherwise. Consistent with the literature, we set $\tau = 1.0$ in our baseline specification. The second-stage 2SLS model is then specified as:

$$\pi_{it} = \alpha_i + \delta_t + \beta_1 s_{it} + \beta_2 \left(s_{it} \times D_{it}^{\theta>\tau}\right) + \phi \text{rel\_p\_lag}_{it} + \zeta_{it}$$

In this model, $\pi_{it}$, $s_{it}$, $\alpha_i$, $\delta_t$, and $\text{rel\_p\_lag}_{it}$ are defined as in Model I. The coefficient $\beta_1$ represents the baseline Phillips curve slope when the labor market is not classified as "tight" ($D_{it}^{\theta>\tau} = 0$). The coefficient $\beta_2$ on the interaction term ($s_{it} \times D_{it}^{\theta>\tau}$) captures the additional change in the slope when the labor market transitions into the "tight" regime ($D_{it}^{\theta>\tau} = 1$). Thus, the effective slope in a tight labor market is $\beta_1 + \beta_2$. A statistically significant and negative $\beta_2$ would provide evidence that the Phillips curve steepens when $\theta_{it} > \tau$. Both $s_{it}$ and the interaction term $s_{it} \times D_{it}^{\theta>\tau}$ are treated as endogenous and are instrumented using $Z_{it}$ ($shift\_share$) and $Z_{it} \times D_{it}^{\theta>\tau}$, respectively. This specification allows us to assess whether the "shape" of the Phillips curve is indeed non-linear and dependent on the prevailing state of labor market tightness.

## 4.3 Method and Inference

All primary specifications are estimated using Two-Stage Least Squares (2SLS). The first stage of our 2SLS procedure regresses the endogenous variable(s) on the chosen instrument(s), all exogenous control variables, and the full set of fixed effects. The predicted values of the endogenous variables from this first stage are then used in the second-stage regression to estimate the parameters of interest. All estimated models incorporate a comprehensive set of fixed effects to control for unobserved heterogeneity. Specifically, we include MSA fixed effects ($\alpha_i$) to absorb time-invariant MSA-specific characteristics that might correlate with both local slack and local inflation. We also include quarter-year time fixed effects ($\delta_t$) to account for common aggregate shocks, national business cycle conditions, secular trends, and, crucially, to difference out national movements in long-run inflation expectations, following the identification strategy of Hazell et al. (2022).

For statistical inference, our baseline results report standard errors clustered at the MSA level. This accounts for potential serial correlation within MSAs over time and arbitrary heteroskedasticity. As a robustness check, particularly considering the potential for spatial correlation across MSAs and serial correlation from overlapping observations in our 4-quarter inflation rates, we will also consider the use of Driscoll and Kraay (1998) standard errors. Finally, given our reliance on an instrumental variable strategy, the relevance of instruments is primarily assessed using first-stage F-statistics. Wu-Hausman tests will also be reported to formally assess the endogeneity of the instrumented regressors.

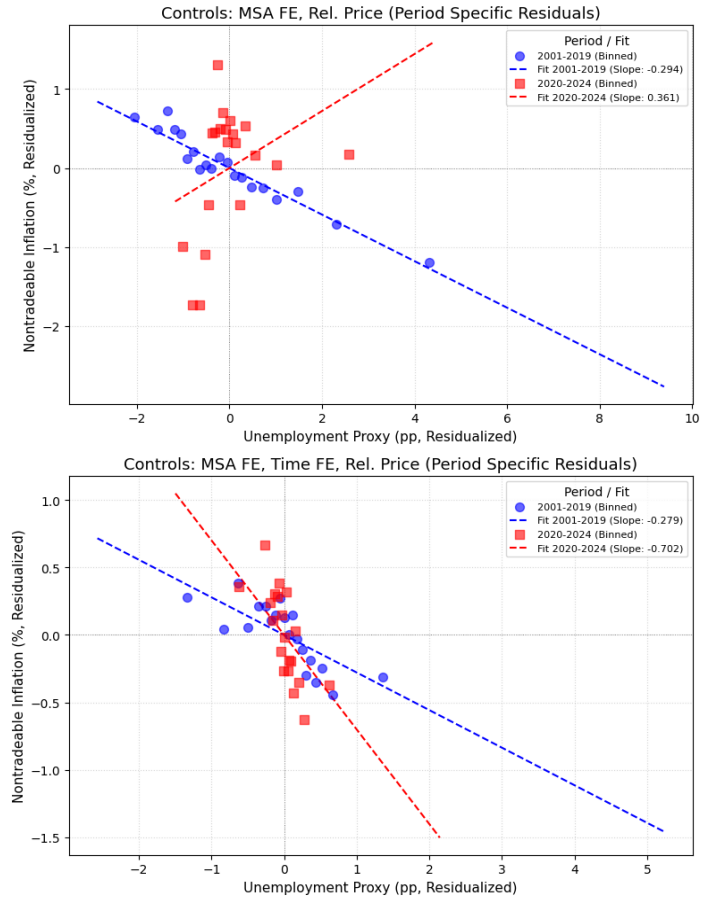

FIGURE IV

Scatterplots: Nontradeable Inflation and Unemployment

## 5. Results and Analysis

### 5.1 The Phillips curve in flux

Our primary goal is to evaluate whether the observed inflation dynamics reflect a fundamental shift in the curve's structural parameters (its "bones") or an activation of its inherent non-linear characteristics (its "shape") under extreme labor market duress. Evidence from Figures II and III already highlighted the unprecedented volatility and heterogeneity in both inflation and labor market conditions since 2020, with aggregate labor market tightness surging past theoretically significant thresholds concurrent with the inflation spike. This backdrop of heightened uncertainty and extreme economic states provides a crucial context for interpreting our regression results.

Key results are displayed in Figure I. Our primary linear specification (Model I) yields a statistically significant ($p < 0.05$) pre-pandemic reduced-form slope $\hat{\psi}_{pre} = -0.7141$. However, the estimated change in this slope during the pandemic period is less significant. These estimates translate into a pre-pandemic structural slope $\hat{\kappa}_{pre} = -0.0779$ and an implied post-pandemic structural slope $\hat{\kappa}_{post} = -0.652$. The result suggests a substantially steeper curve, implying a change in the linear "bones" of the Phillips curve from this model, however, the statistical significance remains a potential confidence concern. The strong first-stage F-statistics (72.5 for slack, 97.1 for the interaction term) and a significant Wu-Hausman test (p=0.0139) lend credibility to the IV strategy and the relevance of the estimated pre-pandemic slope. However, the Within R-squared result suggests that this linear model, even well-identified, somehow cannot capture the whole inflation variations within MSAs after accounting for common shocks and fixed MSA characteristics.

This difficulty in pinning down a stable, significant linear relationship, especially for the pandemic period's change in slope, motivates a closer look at Figure IV (replicated Figure V in Hazell et al. 2022). The "No Time FE" panel of this figure (where variables are residualized against MSA fixed effects and $rel\_p\_lag$ within each period) showed a counter-intuitive positive slope (0.361) for the 2020-2024 period, contrasting with a negative slope (-0.294) for 2001-2019. This anomalous positive slope for the pandemic era in the "No Time FE" specification likely reflects the overwhelming influence of unmodeled aggregate shocks and extreme data volatility during this period. The pandemic brought about substantial, often simultaneous, disruptions to both supply and demand. If these aggregate forces are not adequately controlled for by time fixed effects, the estimated relationship between local slack and local inflation can be severely distorted. The "No Time FE" plot's post-pandemic data points also appeared highly dispersed, suggesting that outliers or extreme heterogeneity across MSAs in their response to these unique shocks might be driving this OLS fit. The fact that the slope turns negative and steeper (-0.702 for 2020-2024 vs. -0.279 for 2001-2019) in the "With Time FE" panel of Figure V underscores the critical role of time fixed effects in absorbing these pervasive national-level influences and revealing a more theoretically consistent underlying Phillips curve.

Together, these results point towards the potential importance of non-linearities, where the slope itself is a function of the economic state, an idea we explore with Model II. While our analysis, like much of the literature using regional data, effectively controls for national inflation expectations via time fixed effects (Hazell et al., 2022), the nature of the pandemic shocks may necessitate exploring these non-linear "shapes" more directly. Our findings on the linear "bones" suggest either remarkable stability in the face of unprecedented shocks or a relationship obscured by the sheer magnitude and peculiarity of pandemic-era data that linear models struggle to parse.

## 5.2 Unmasking the "Shape"

As shown in Table I, Model II estimates a baseline slope in non-tight labor markets ($\hat{\beta}_1 = -0.7505$) that is quantitatively very close to the pre-pandemic slope found in Model I ($\hat{\psi}_{pre} = -0.7141$). This consistency across models for the "normal" or "slack" labor market regime is reassuring, suggesting a relatively stable, conventional Phillips curve relationship when the labor market is not under extreme duress. Both estimates indicate that a one-unit increase in our $v/u$-based slack measure is associated with roughly a 0.71-0.75 percentage point decrease in non-tradable inflation. This forms a consistent estimate of the Phillips curve's slope on its flatter segment.

TABLE I

Comparative Results of Phillips Curve Estimations

|  | Model I | Model II |
|---|---|---|
| Dependent Variable: pi_core_4q | Interaction with Pandemic | Interaction with Market Tightness |
| **Key Slope Estimates** | | |
| slack (Pre-Pandemic / Non-Tight Market) | -0.7141** | -0.7505** |
|  | (0.3360) | (0.3361) |
| slack×pandemic_period_num ($\Delta\psi$) | -0.3288 | |
|  | (0.7423) | |
| slack×tight_market_dummy_m2 ($\beta_2$) |  | -0.2545 |
|  |  | (0.5855) |
| **Implied Slope** | | |
| slack (Post-Pandemic / Tight Market) | -1.0429 | -1.0050 |
| **Structural Slope Estimates** ($\hat{\kappa}$) | | |
| Pre-Pandemic / Non-Tight Market | -0.0779 | -0.0828 |
| Post-Pandemic / Tight Market | -0.6520 | -0.6286 |

The central question for Model II, however, is whether this slope changes when the labor market becomes "tight" ($\theta_{it} > 1$). The estimated additional impact on the slope in such tight conditions $\hat{\beta}_2$ is -0.2545. This point estimate implies that the Phillips curve does indeed steepen further (to an implied total slope of -1.0050) when vacancies outnumber the unemployed, a finding directionally consistent with Benigno and Eggertsson's (2023) framework. The magnitude of this implied tight-market slope (-1.0050) is also strikingly similar to the implied post-pandemic slope from Model I (-1.0429). This similarity suggests that the apparent steepening observed in linear models when considering the entire pandemic period might be driven by the economy more frequently operating in this theoretically steeper, tight-labor-market portion of a non-linear Phillips curve, rather than a uniform shift in the linear structure itself. The pandemic era was unique in pushing $\theta_{it}$ to historically high levels, potentially activating this non-linear "shape."

While the "shape" hypothesis is appealing and the data hints at its relevance, our MSA-level analysis with the current specification and instrument does not provide definitive statistical confirmation of this non-linear steepening. Several factors could contribute to this imprecision. The "tight labor market" regime ($\theta > 1$), while experienced more frequently post-2020, still constitutes a smaller portion of our overall sample compared to "normal" or "slack" periods. Estimating a distinct slope for this regime with high precision can be challenging, especially given the substantial heterogeneity across MSAs in their labor market dynamics during the pandemic (as highlighted by Figure III). Furthermore, the simple binary threshold might be an oversimplification of a more continuous, smooth transition, or the true threshold might vary across

MSAs or over time. Benigno and Eggertsson (2023) themselves acknowledge that their theoretical threshold $\theta^*$ can vary and use $\theta^* = 1$ as a "reasonable approximation". The extreme supply shocks and unprecedented policy interventions during the pandemic also created a extremely noisy and unpredictable environment, making it difficult to isolate the impact of labor market tightness on inflation, even with strong instruments. The fact that the cost of hiring new workers, rather than average wages, is posited to be the key driver of marginal costs in a labor shortage also presents an empirical challenge, as our aggregate MSA-level slack measure might not perfectly capture this specific channel.

# 6. Conclusion

The COVID-19 pandemic and its aftermath have presented an unprecedented economic landscape, reigniting critical debates about the nature and stability of the Phillips curve. This paper has sought to contribute to this discourse by examining the U.S. Phillips curve through the lens of Metropolitan Statistical Area (MSA) level data from 2001 to 2024, distinguishing between potential changes in its underlying structural slope ("bones," $\kappa$) and state-dependent variations in its effective responsiveness contingent on labor market tightness ("shape," $\lambda(\theta)$).

Our analysis of the linear "bones" of the Phillips curve (Model I) indicates a statistically significant, conventional negative slope between non-tradable inflation and labor market slack in the pre-pandemic period. However, we do not find robust statistical evidence of a uniform, linear shift in this slope during the pandemic and post-pandemic era. While point estimates suggested a potentially steeper curve, this change was not statistically distinguishable, implying that the fundamental linear structure of the inflation-slack trade-off, when averaged across all states of the economy, may not have been permanently altered by the pandemic in a statistically identifiable way with linear specification. The baseline slope in non-tight labor markets of the non-linear "shape" investigation (Model II) was found to be negative and significant, consistent with Model 1's findings.

Several limitations to our analysis warrant discussion and point towards avenues for future research. First, while our use of MSA-level data and a shift-share instrument addresses common identification challenges related to aggregate inflation expectations and endogeneity, the primary challenge appears to be empirically pinning down state-dependency with high precision using regional data during such an extraordinary and volatile period as the pandemic. Second, our chosen measures for labor market slack and the proxy for relative non-tradable prices might not perfectly capture the relevant economic constructs across all economic states, particularly during the pandemic, when traditional relationships may have been distorted. For instance, the reliance on lagged slack can be problematic when market conditions change rapidly. Similarly, the behavior of our $rp\_reg\_proxy$ might have been unusual post-2020. While our main regression models use a contemporaneous slack measure, its interaction with state or period dummies might still be affected by the unique dynamics of its components (v and u). Future research could explore alternative, potentially time-varying, measures of slack or more sophisticated controls for relative price movements. Third, the non-linear model employed (a simple threshold interaction) is a specific and relatively constrained representation of the state-dependency theorized by Benigno and Eggertsson (2023, 2024). While their framework implies a piecewise linear Phillips curve in logs, the exact threshold ($\theta^*$) may vary over time or across regions, and the transition might be smoother. More granular data, perhaps at the firm level, or more sophisticated techniques for estimating non-linear panel data models (e.g., smooth-transition regression, grid search, etc.) could provide more precise estimates of these "shapes."

To conclude, our research underscores that understanding the Phillips curve requires a multi-faceted approach. It calls for continued investigation into the precise nature of non-linearities, the role of rapidly evolving labor market structures (such as changes in matching efficiency or the share of unattached workers), and the development of more robust empirical strategies to disentangle structural parameters from state-contingent effects during periods of profound economic upheaval. The exploration to understand whether the recent inflation surge was a temporary aberration driven by unique shocks amplified by a non-linear curve, or a harbinger of a more fundamental shift in inflation dynamics, remains a critical task.